\documentclass[nofootinbib,twocolumn]{revtex4-1}
\usepackage{graphicx}
\usepackage{amsmath}
\usepackage{amssymb,bm}

\usepackage{color}
\usepackage{xcolor}
\usepackage{verbatim}

\definecolor{MS-color}{RGB}{128,0,128}

\begin{document}

\title{Field dependence of the vortex-core sizes in dirty two-band superconductors}

 \author{A. Vargunin}
 \affiliation{Department of
Physics and Nanoscience Center, University of Jyv\"askyl\"a, P.O.
Box 35 (YFL), FI-40014 University of Jyv\"askyl\"a, Finland}
\affiliation{Institute of Physics, University of Tartu, Tartu, EE-50411, Estonia}
 \author{M. A.~Silaev}
 \affiliation{Department of
Physics and Nanoscience Center, University of Jyv\"askyl\"a, P.O.
Box 35 (YFL), FI-40014 University of Jyv\"askyl\"a, Finland}
\date{\today}

\begin{abstract}
We study the structure of Abrikosov vortices in two-band superconductors for different external magnetic fields and different parameters of the bands. The vortex core size determined by the coherence lengths are found to have qualitatively different behaviour from that determined by the quasiparticle density of states spatial variation. These different vortex core length scales  coincide near the upper critical field, while the discrepancy between them becomes quite significant  at lower fields.    Within the diffusive approximation we demonstrate 
{several generic regimes in the field dependence of the vortex core sizes}
determined by the disparity of diffusion constants in the two bands.  
\end{abstract}

\maketitle

\section{Introduction}
Vortex matter in multiband superconductors demonstrates many unusual properties which are drastically different from that in single-band materials\cite{PhysRevB.81.020506,
PhysRevB.85.094511,PhysRevLett.102.117001,eskilsen2002,PhysRevLett.95.097004, 
PhysRevB.81.214501,Fente2016}. The origin of non-trivial new effects comes from the greatly enhanced number of the available degrees of freedom in the system consisting of Cooper pairs and quasiparticles residing in several different bands. In this case the condensates in general tend to have different coherence lengths\cite{PhysRevLett.105.067003,
PhysRevB.72.180502, PhysRevLett.105.067003} sharing the same critical temperature and the single divergent scale near $T_c$ \cite{PhysRevB.84.094515} . With increasing the coupling between condensates their length scales become essentially the same\cite{PhysRevB.85.134514}.  Although being quite important characteristics the coherence lengths are not directly measurable. For example the sizes of Abrikosov vortices in the two-band superconductor MgB$_2$ measured with the scanning tunnelling microscopy (STM) local probes  appear to be significantly different from the coherence length inferred from the upper critical field\cite{eskilsen2002}. This physics is explained by the different localization scales of local density of states profiles in different bands determined by the disparity of diffusion coefficients\cite{PhysRevLett.90.177002}. 


The high resolution of STM allows to explore individual vortex cores in details by measuring the quasiparticle local density of  states (LDOS) 
\cite{PhysRevLett.64.2711, PhysRevLett.62.214,GH,suderow2014, Fente2016}. The LDOS profiles $N(r)$ are  essentially determined by the spatial order parameter distribution $\Delta(r)$ near the vortex core. However the vortex core size determined from the STM tunnelling conductance  depends on the temperature and bias \cite{volodin1997} indicating the spatial and energy variation of the LDOS of localized quasiparticle states trapped close to the vortex center. 

The relation between {zero-energy LDOS} $N(r)$ and $\Delta(r)$ is quite straightforward in diffusive superconductors when the magnetic field $B$ is close to the upper critical field $H_{c2}$ so that $H_{c2}-B \ll H_{c2}$. 
In this regime as shown by de Gennes\cite{dG} the following relation holds
 \begin{equation} \label{Eq:deGennes}
     N(r)= 1 -  2|\Delta (r)|^2/\Delta_0^2
 \end{equation}
 where $N$ is normalized to the normal metal DOS and $\Delta_0$ is the gap function amplitude in the absence of magnetic field. 
At lower fields the relation between $N(r)$ and $\Delta (r)$ has not been  checked even in the simplest case of single-band superconductors. 
 In the present paper we demonstrate that in general the behaviour of these two profiles with decreasing magnetic field becomes quite different so that it is not possible to extract the information about coherence length from STM measurements by applying directly the Eq.(\ref{Eq:deGennes}).  
 
The behaviour of gap and LDOS profiles can be even more intriguing in two-band superconductors. According to the recent experiments \cite{Fente2016} the vortex sizes measured by STM in 2H-NbSe$_2$ and 
2H-NbS$_2$ compounds demonstrate much weaker magnetic field dependencies than in the single-band materials. 
Interpreting these data using de Gennes relation \cite{dG} results in the conclusion about the mostly field-independent condensate length scales in the two-band superconductors. 
Here we report the results of exact numerical calculations in the framework of the multiband Usadel theory. 
We find that  in the two-band superconductor the vortex core sizes $w_{\Delta_1}$, $w_{\Delta_2}$ determined by the gap function profiles  in different bands $\Delta_{1,2}(r)$ in general have no distinct correlation with the widths $w_{\sigma_1}$, $w_{\sigma_2}$ of the corresponding LDOS distributions $N_{1,2}(r)$. We illustrate that 
for the distinct disparity between diffusion coefficients in different bands, the vortex core sizes $w_{\sigma_1,\sigma_2}$ and $w_{\Delta_1,\Delta_2}$
can show qualitatively different behaviour as functions of the magnetic field. 
For the large enough interband pairing the gap function distributions  $\Delta_{1}(r)$
and $\Delta_{2}(r)$ are  mostly identical so that $w_{\Delta_1}\approx w_{\Delta_2}$. However the profiles of $N_1(r)$
and $N_{2}(r)$ are strongly different except of the high field regime when the modified de-Gennes relation restores and all length scales coincide.  
We demonstrate that in the superconducting band with the smallest diffusion coefficient the zero-energy LDOS length-scale shows quite weak magnetic field dependence in accordance with STM data in multiband superconductors\cite{Fente2016}.
At the same time the width of LDOS profile in the band with larger diffusion coefficient grows with decreasing magnetic field in a rate which is typically faster than the growth of healing  lengths $w_{\Delta_1}$, $w_{\Delta_2}$ characterizing the order parameter distributions. 

 The structure of this paper is as follows. In Sec. \ref{theory} we introduce the formalism of the quasiclassical Green's functions to describe the properties of dirty multiband superconductors and discuss numerical approach for the solution of self-consistency problem. In Sec. \ref{results}, the results of the numerical calculations are presented. First, in Sec. \ref{resultsA} we checked method in the single-band limit. In Sec \ref{resultsB}, we examine in details the field dependencies of the gap and LDOS profiles in different bands and calculate characteristic length scales in two-band model. The work summary is given in Sec. \ref{summary}.


\section{Model}\label{theory}
We use the formalism of quasiclassical Green's functions (GF) and introduce retarded/advanced GF, $\hat g^{R/A}_k$, for two-band ($k=1,2$) superconductor which obey in the diffusive limit the Usadel equation
 \begin{equation}\label{spectraleq}
 D_k\hat\nabla ( \hat g^{R/A}_k\cdot\hat \nabla\hat g^{R/A}_k) + [i\varepsilon\hat\tau_3+i\hat\Delta_k, \hat g^{R/A}_k ]=0.
   \end{equation} 
Here $D_k$ is diffusion constant in each band, $\hat \Delta_k=\left(\begin{array}{cc} 0 &  \Delta_k \\-\Delta^\ast_k &  0
\end{array}\right)$ is the gap operator, and $\hat\nabla=\nabla-i\pi\phi_0^{-1}{\bm A}[\hat\tau_3,]$, where $\hat\tau_3$ is the Pauli matrix, square brackets denote commutator operation
 and $\phi_0=\pi /e$ is the flux quantum. Note that we use theoretical units $k_B=\hbar=c=1$.

To describe the vortex structure at arbitrary fields we employ the circular cell approximation \cite{ihle1971b,WattsTobin1974,Rammer1987,Rammer1988}. Within this approach the unit cell of the hexagonal vortex lattice hosting a single vortex is replaced 
by a circular cell with the centre at the point of superconducting phase singularity. 
Inside circular cell, the gap and magnetic field distributions are taken radially symmetric with respect to the cell centre. Below we consider the vortex state in the limit of large values of the Ginzburg-Landau parameter, $\kappa\gg1$. In this case, magnetic field $B$ is constant inside circular cell and the vector potential can be taken in the form ${\bm A}(r)={\bm \varphi}Br/2$.
The periodicity of the 
lattice solution is modelled by the special choice of the 
boundary conditions, namely the vanishing the supercurrent density at the circular-cell boundary. At that, the circular-cell radius is uniquely defined by magnetic induction, $R=\sqrt{\phi_0/(\pi B)}$ so that there is exactly one flux quantum $\phi_0$ passing through the unit vortex cell. 

In the $\theta$-parameterization, GF in Nambu space read as
\begin{align}
\hat g^R_k=
\left( \begin{array}{cc}
\cosh\theta^{(k)} & 
\sinh\theta^{(k)} e^{i\varphi_k}
\\
-\sinh\theta^{(k)} e^{-i\varphi_k} & -\cosh\theta^{(k)}\end{array}\right),
\end{align}
where $\varphi_k$ is band-gap phase. In cylindrical coordinates, Eq. (\ref{spectraleq}) can be rewritten for complex angles $\theta^{(k)}$ as
 \begin{align}\label{theta}
&D_kr\partial_r(r\partial_r\theta^{(k)})-D_k(1-r^2/R^2)^2\sinh\theta^{(k)}\cosh\theta^{(k)}\nonumber\\
&+2ir^2(\varepsilon\sinh\theta^{(k)}-|\Delta_k|\cosh\theta^{(k)})=0.
 \end{align}
This set of equations has to be solved self-consistently with gap order parameters determined by conditions
\begin{align}\label{gaps}
|\Delta_k|=2\pi T\sum_{k^\prime}\lambda_{kk^\prime}\sum_{\omega_n>0}\sin \theta_n^{(k^\prime)},
\end{align}
where $\lambda_{kk^\prime}$ are intra- and interband interaction constants which form matrix $\hat \lambda$, $\omega_n=\pi T(2n+1)$ are Matsubara frequencies and Matsubara GF parametrized by $\theta_n^{(k)}$ satisfy (\ref{theta}) after substitution $\theta^{(k)} \to-i\theta_n^{(k)}$ and $\varepsilon\to i\omega_n$. At that, boundary conditions read as $\theta_n^{(k)}(r=0)=0$ and $\partial_r\theta_n^{(k)}(r=R)=0$ leading to zero gradient of gap modulus at the vortex-cell boundary.

We normalize magnetic field by upper critical one which in the two-band model is determined by condition $|\hat A|=0$, where $|\hat A|={\rm Det}\hat A$ and
 \begin{align}
A_{kk^\prime}= (\hat\lambda^{-1})_{kk^\prime}+\delta_{kk^\prime}\left[ f_k(T)-G_0+\ln (T/T_c) \right].
 \end{align}
 Here 
 $f_k=\Psi\left[1/2+q_k/(2\pi T)\right]-\Psi(1/2)$, $\Psi$ is digamma function, $q_k=eD_kH_{c2}$, 
$2G_0=({\rm Tr}\hat\lambda-\lambda_0)/|\hat\lambda|$ and $\lambda_0^2=({\rm Tr}\hat\lambda)^2-4|\hat\lambda|$. Except for zero and critical temperatures, $H_{c2}(T)$ has to be calculated numerically. In the limit $T\to 0$, upper critical field is given by the expression $2\sqrt{q_1q_2}/T_c=\pi e^{g_+/2-C}$, where\cite{PhysRevB.67.184515} 
\begin{align}
&(g_++\lambda_0/|\hat\lambda|)^2=
\left[ \ln (D_1/D_2) - (\lambda_{11}-\lambda_{22})/|\hat\lambda|\right]^2
\nonumber\\
&+4(\lambda_{11}\lambda_{22}-|\hat\lambda|)/|\hat\lambda|^2,
\end{align}
and $C\approx0.577$ is the Euler constant. Note that in weak-coupling limit 
the critical temperature of a two-band superconductor is $T_c=\omega_D/(2\pi \Omega)$, where $\omega_D$ is Debye energy cut-off and $4\Omega=e^{G_0-C}$.

To find the self-consistent order parameter distributions we start by calculating gap order parameters in the cell taking first initial distributions of $|\Delta_{1,2}(r)|$. By initializing guess functions $\theta_{n}^{(1,2)}$ for each $n$, we linearise equation for Matsubara GF around $\theta_{n}^{(1,2)}$ and solve the linear problem numerically by apply sweeping method. Solution provides correction to $\theta_{n}^{(1,2)}$ and refined guess function is used for next iteration. By performing sufficient number of iterations, procedure converge to the Matsubara GF which are substituted into right-hand side of Eq. (\ref{gaps}) to obtain correction to the initial gap functions $|\Delta_{1,2}|$. By applying refined gap functions, we repeat scheme from the beginning and find gaps in iterative process with needed precision.

The zero-energy LDOS in different bands 
is given in $\theta$-parametrization by 
$N_k=\cos ({\rm Im}\theta^{(k)})$ at $\varepsilon=0$ . To find $N_k$, we consider imaginary part of Eq. (\ref{theta}) at zero energy with gap profiles found beforehand. We solve it numerically by starting from guess distributions for $\theta^{(k)}$. We linearise (\ref{theta}) around $\theta^{(k)}$ and solve the linear problem numerically by sweeping method. Solution gives correction to $\theta_{k}$ which is used to construct refined guess distribution and employ iteration procedure.

\section{results}\label{results}
\subsection{Single-band limit}\label{resultsA}
The approach presented in Sec. \ref{theory} reduces to the single-band model, if $\lambda_{12}=\lambda_{21}=0$ and $D_{1,2}=D$. For $\lambda_{11}>\lambda_{22}$, it corresponds to the description of the independent stronger-superconductivity band.
%
  \begin{figure}[t!]
  \includegraphics[width=0.99\linewidth]{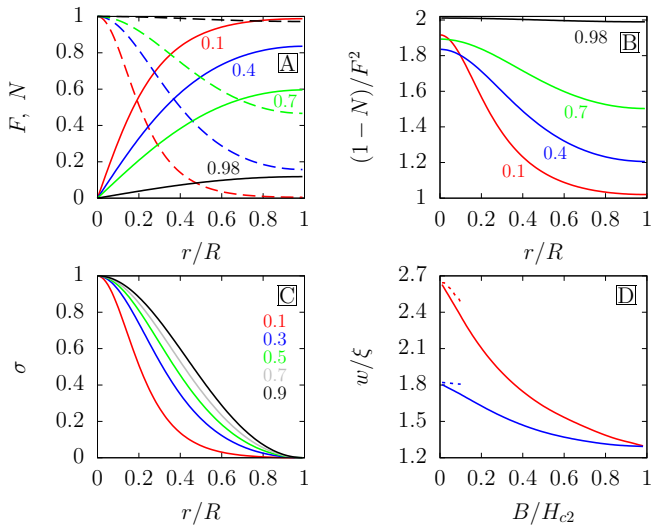}
  \caption{\label{f1} 
  (Color online) Vortex structure in the single-band model at $T/T_c=0.05$. (A) Normalized gap distribution (solid) inside vortex cell, $F=|\Delta|/\Delta_0$. Numbers near each curve indicate the value of $B/H_{c2}$. The dashed curves with the same color show zero-energy LDOS $N$. (B) Proportionality coefficient between $1-N$ and $F^2$ within vortex cell for different magnetic fields. Note that we obtain 
  $(1-N)/F^2=2$ near $H_{c2}$ in agreement with Eq. (\ref{Eq:deGennes}). (C) LDOS variation $\sigma=\delta N(r)/\delta N(0)$ for different ratios $B/H_{c2}$. (D) The field behaviour of the vortex-core size $w=w_\Delta$ determined by the half-width of squared gap $|\Delta|^2$ (red) and the one $w=w_\sigma$ defined by the half-width of LDOS variation $\sigma$ (blue). Both quantities are shown in units of  $\xi=\sqrt{D/(2\pi T_c)}$.
  The dashed curves are calculated by means of Eq. (\ref{spectraleq}) with substitution $\hat\nabla\to\nabla$, see discussion in the text.  	 }
  \end{figure}
Fig. \ref{f1} demonstrates the results of the self-consistent numerical calculations for single-band superconductor. As magnetic field increases, the radius of circular vortex cell reduces resulting in the suppression of the maximal gap value achieved at the cell boundary, see Fig. \ref{f1}A. At that, inhomogeneity of zero-energy LDOS $N$ inside vortex cell smooths out by rising field.
 
 In Fig. \ref{f1}C we plot LDOS variation $\sigma=\delta N(r)/\delta N(0)$, where $\delta N(r)=N(r)-N(R)$, for different magnetic fields. In single-band limit, these curves do not depend on material parameters such as the diffusion coefficient $D$ and $T_c$. 
 From definition it is clear, that LDOS variation $\sigma$ is characterized by the same half-width as LDOS $N$ itself.

To check our numerical results we test the obtained profiles against the validity of de Gennes relation (\ref{Eq:deGennes}) at low temperatures and close to $H_{c2}$. 
In Fig. \ref{f1}B we show the ratio $(1-N (r))/F^2(r)$, where $F(r)=|\Delta(r)|/\Delta_0$,
for different magnetic fields at temperature $T=0.05 T_c$. 
For high fields this ratio is constant in agreement with Eq. (\ref{Eq:deGennes}).
However, for the lower fields, $B/H_{c2}\lesssim 0.5$, it is significantly inhomogeneous meaning that LDOS evolution inside vortex cell is essentially different from the order parameter. As a result, LDOS measurements for sparse vortex lattices in general cannot be used to quantify the length scale of the superconducting order parameter.

Fig. \ref{f1}D demonstrates the field dependencies for half-widths $w_\Delta$ and $w_\sigma$ of squared gap $|\Delta|^2$ and LDOS variation  $\sigma$, respectively. Two half-widths shown in Fig. \ref{f1}D overlap in the limit $B\to H_{c2}$, where spatial profiles of LDOS and $|\Delta|^2$ become identical, see black curve in Fig. \ref{f1}B. In this case, we expect that Abrikosov vortex lattice solution governs the behaviour of the gap order parameter so that half-width is determined by the size of the superconducting nucleus. By using known analytic solution for the gap at $H_{c2}$\cite{GH} given by $F(r)\propto re^{-r^2/(2R^2)}$, 
we obtain $w_\Delta \approx 0.48R$. At low temperatures, upper critical field is determined by $q/T_c=\pi/(2e^C)$ so that $w_\Delta\approx 1.3\xi$, where $\xi^2=D/(2\pi T_c)$, in agreement with numerical value presented in Fig. \ref{f1}D.

For lower fields, half-widths $w_\sigma$ and $w_\Delta$ have qualitatively different behaviours manifesting significant difference between the squared gap and LDOS profiles and violation of simple relation (\ref{Eq:deGennes}). The half-width found for the squared gap coincides with previous calculations \cite{GH} and scales approximately as $w_\Delta \sim(B/H_{c2})^{-1/3}$ in the intermediate fields. At the same time, the half-width of LDOS $w_\sigma$ changes with the field slower than that.

For very sparse vortex lattices, $B/H_{c2}\ll1$, both scales are characterized by linear field-dependence and for the gap we obtain $w_\Delta(B)/w_\Delta(0)\simeq1-B/H_{c2}$. Such a behaviour indicates that spatial evolution of the gap profile is affected by the term linear in the vector potential. This behaviour can be checked by calculating vortex core sizes in the absence of vector potential. In result instead of the linear behaviour we get the  low-field plateaus in the dependencies of 
$w_\Delta(B)$ and $w_\sigma(B)$ shown by the dashed lines in Fig. \ref{f1}D. Thus the absence of any pronounced variation of vortex core sizes at small magnetic fields found by STM experiments\cite{Fente2016} cannot be attributed to the specific range of magnetic field $B\ll H_{c2}$ studied there. 
On the contrary as we demonstrate below, almost field-independent vortex core sizes can be naturally obtained within the minimal two-band model of superconducting state. 
%

  \begin{figure}[t!]
  \includegraphics[width=0.99\linewidth]{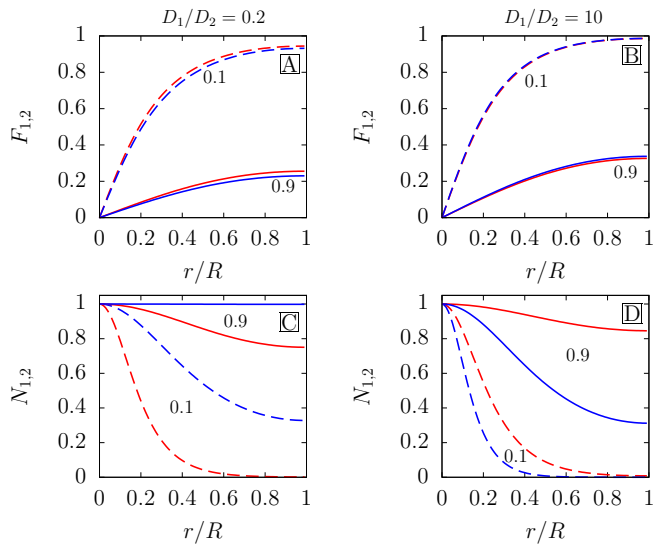}
  \caption{\label{f2} 
  (Color online) Vortex structure in the two-band model at $T/T_c=0.1$ for $D_1/D_2=0.2;\ 10$ (left and right column, respectively). (A,B) 
  Gap profiles in each band normalized by bulk value, $F_k=|\Delta_k|/\Delta_{k0}$. Dashed red/blue curves correspond to $F_{1,2}$ at $B/H_{c2}=0.1$ and pair of solid curves to $B/H_{c2}=0.9$. (C,D) LDOS in each band $N_k$ for small (dashed) and high (solid) fields. Red/blue colours correspond to $N_{1,2}$. }
  \end{figure}
\subsection{Two-band model}\label{resultsB}
The two-band superconductivity is defined by the matrix of interaction constants $\hat\lambda$ and by the ratio of diffusion coefficients in the bands $D_1/D_2$. For calculations we consider typical parameters \cite{MgB2}, namely, $\lambda_{11}=0.1012$, $\lambda_{12}=0.0336$, $\lambda_{21}=0.0264$ and $\lambda_{22}=0.0448$, and consider evolution as $D_1/D_2$ changes.

Fig. \ref{f2} shows that gap function  profiles in different bands look very similar. If one normalizes gaps by their maximal value reached at cell boundary then the difference between normalized gap distributions practically vanishes.This result does not depend on the values of $D_{1,2}$ despite that these
parameters define coherence lengths in the absence of Josephson coupling between the bands. 
In the considered case of sufficiently strong interband interaction, the mixing between superconducting condensates of separate bands is so efficient that healing length of different gap functions $\Delta_{1,2}$ become almost identical. 

In contrast to the gap profiles, LDOS in separate band is strongly affected by the band diffusion coefficients. This is seen from Usadel Eq. (\ref{theta}) where 
 characteristic lengths of solutions $\theta^{(1,2)}(r)$ differ by the factor  $\sqrt{D_1/D_{2}}$. Fig. \ref{f2}C,D confirms this behaviour showing that LDOS in the band with smaller diffusion coefficient changes at shorter distances than 
the one in the band with larger diffusion coefficient.

Apart from the characteristic scales determined by the diffusion coefficients   there is another characteristic length which is the circular cell radius $R$. 
%
%
Changing the diffusion coefficients in different bands independently one can obtain the  unusual situation peculiar for two-band model when the length scale of LDOS variation in one of the bands is much larger than the  cell radius. In this case the LDOS  corresponding to the band with larger diffusion coefficient changes  
within vortex cell very weakly in the wide range of the fields. This situation is illustrated in Fig. \ref{f2}C where the band with weaker superconductivity, $\Delta_2<\Delta_1$, 
has larger diffusion coefficient $D_1/D_2=0.2$. The LDOS in this band  (blue curves in Fig. \ref{f2}C) changes within vortex cell very weakly in the wide range of the fields (already for $B/H_{c2}\gtrsim 0.3$). Thus, in this case the characteristic length scale for $N_2$ variation, $w_{\sigma_2}$, is expected to scale with cell radius $R\propto 1/\sqrt{B}$. As we see below, this is indeed the case. 

  \begin{figure}[t!]
  \includegraphics[width=0.99\linewidth]{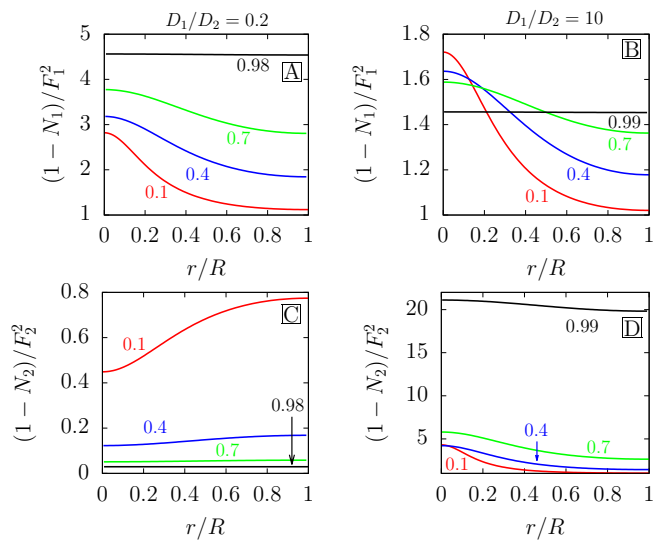}
  \caption{\label{f3} 
  (Color online) Spatial variation of proportionality coefficient between $1-N_k$ and $F_k^2$ in two-band model with $D_1/D_2=0.2;\ 10$ (left and right column, respectively). (A,B) Coefficient for the stronger-superconductivity band as $B/H_{c2}$ indicated by the colour numbers increases. (C,D) Coefficient for weaker-superconductivity band. }
  \end{figure}

Another unusual situation generic for two-band model only 
can be realized when LDOS corresponding to the band with smaller diffusion coefficient varies within vortex cell on a distance which is much smaller than cell radius. This case is demonstrated in Fig. \ref{f2}D where band with weaker gap has smaller diffusion coefficient, $D_1/D_2=10$, and variations of its
LDOS (blue curves in Fig. \ref{f2}D)
are only weakly affected by the changes 
of the vortex cell radius under magnetic field. As we see below, this results in the weak field dependence of the length scale related to the LDOS variations of the relevant band.

  \begin{figure}[t!]
  \includegraphics[width=0.99\linewidth]{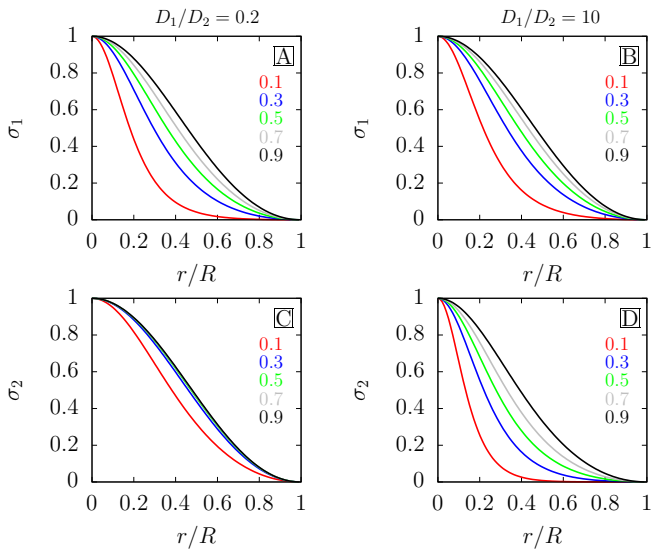}
  \caption{\label{f4} 
  (Color online) LDOS variations $\sigma_k=\delta N_k(r)/\delta N_k(0)$ inside the vortex as the value of normalized magnetic field $B/H_{c2}$ indicated by different colours increases. Left and right column correspond to the cases $D_1/D_2=0.2;\ 10$, respectively.}
  \end{figure}

Let us discuss the relations between  gap functions $\Delta_{1,2}(r)$ and   LDOS deviations from the normal state, $1-N_{1,2} (r)$, shown Fig. \ref{f3} for the 
two-band model. One can see that analogously to the single band case  these profiles coincide only for the high fields close to  the upper critical one.  This limit can be approached analytically. For $B\approx H_{c2}$, the order parameter is small and it can be written as $|\Delta_k|=c_k\Delta_{k0}r e^{-r^2/(2R^2)}$, where $c_k$ is small constant, see \cite{golubov1988}. The zero-energy solution of spectral Eq. (\ref{theta}) is then given by $\mathrm{Im}\theta_k=\alpha_k|\Delta_k|$, where $\alpha_k\sqrt{q_1q_2} B/H_{c2}=-\sqrt{D_{3-k}/D_k}$. 
As a result, we obtain the relation between the LDOS and order parameter in the two-band model 
which is valid at fields very close to upper critical one
\begin{align}\label{propc}
& N_k = 1-\frac{|\Delta_k|^2}{2e^2 D^2_k H^2_{c2} }.
\end{align}
This formula generalizes the de Gennes 
 relation (\ref{Eq:deGennes}) for the multiband system and arbitrary temperatures. Indeed, in the one band case one restores the relation (\ref{Eq:deGennes}) at low temperatures $T\to 0$ by taking into account single-band limiting value $eDH_{c2}/\Delta_0=1/2$.
However, in two-band case the relation between bulk gap in particular band and the upper critical field depends strongly on two-band model parameters, in particular, the ratio of diffusion constants. For our parameters of two-band model and $T=0.1T_c$ we have $\Delta_{1,20}/T_c\approx 2.05;0.81$ and $\sqrt{q_1q_2}/T_c\approx 1.51;0.38$ for $D_1/D_2=0.2;10$, respectively. According to the Eq. (\ref{propc}), the proportionality coefficient between $1-N_k$ and $|\Delta_k|^2/\Delta_{k0}^2$ is then given by $4.56;1.45;0.03;23$ for the cases shown in Fig. {\ref{f3}}A,B,C,D, respectively. These values coincide with the black curves in Fig. \ref{f3} remarkably well.

Next we calculated LDOS variations within the vortex cell defined as $\sigma_k=\delta N_k(r)/\delta N_k(0)$, where $\delta N_k(r)=N_k(r)-N_k(R)$, shown in Fig. \ref{f4}. Contrary to the single-band model where $\sigma$ has universal field behaviour, two unusual regimes can be realized in two-band superconductor depending on the value of $D_1/D_2$ parameter.

 The case $D_1\ll D_2$ shown in Fig. \ref{f4}C is characterized by the leading role of the vortex-cell radius $R$ in the spatial variations of LDOS in the band with larger diffusion constant, see discussion of Fig. \ref{f2}C. As a result, the field dependence of $\sigma_2$ is governed by $R$ so that magnetic field modifies $\sigma_2(r/R)$ profiles extremely weakly, see Fig. \ref{f4}C.
Recent STM measurements of multiband systems $\beta-$Bi$_2$Pd\cite{Bi2Pd}, 2H-NbSe$_{1.8}$S$_{0.2}$ and 2H-NbS$_2$ demonstrate very similar behaviour \cite{Fente2016} suggesting that these compounds have large  disparity between the diffusion coefficients in different bands.
 
The opposite regime $D_1\gg D_2$ illustrated in Fig. \ref{f4}D is described by weak field-dependence of LDOS profiles in the band with smaller diffusion coefficient, see discussion of Fig. \ref{f2}D. This results in the very diverse field modifications of $\sigma_2(r/R)$ curves, see Fig. \ref{f4}D, which can be also used as a fingerprint of multiband superconductivity.

   \begin{figure}[t!]
   \includegraphics[width=0.99\linewidth]{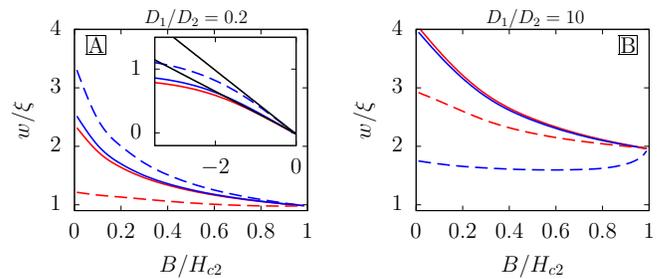}
   \caption{\label{f5} 
    (Color online) Field dependence of the length scales  normalized to $\xi=\sqrt{\sqrt{D_1D_2}/(2\pi T_c)}$ for $D_1/D_2=0.2;10$. Red/blue solid lines correspond to $w=w_{\Delta_{1,2}}$ determined as the half-widths of $|\Delta|_{1,2}^2$ and red/blue dashed lines to $w=w_{\sigma_{1,2}}$ defined as the half-widths of $\sigma_{1,2}$, respectively. The inset in panel A is the plot of $\log w/\xi$ \textit{vs} $\log B/H_{c2}$ for $w_{\Delta_{1,2}}$ (red/blue solid) and $w_{\sigma_2}$ (blue dashed). Linear dependencies $y=-x/2+\mathrm{const}$ (upper black) and $y=-x/3+\mathrm{const}$ (lower black) indicate the scaling $w_{\sigma_2}\sim(B/H_{c2})^{-1/2}$ and $w_{\Delta_{1,2}}\sim(B/H_{c2})^{-1/3}$ in the vicinity of $H_{c2}$.} 
   \end{figure}
Finally, we have calculated field dependencies for the lengths $w_{\Delta_{1,2}}$ determined as the half-widths of $|\Delta|_{1,2}^2$ and $w_{\sigma_{1,2}}$ defined as the half-widths of $\sigma_{1,2}$,
see Fig. \ref{f5}. At higher fields, all length scales approach same value which differs from the one obtained in single-band limit. According to analytical solution for superconducting nucleus, the half-width of squared gap at $H_{c2}$ is given by $w\approx 0.48R$. By using values $\sqrt{q_1q_2}/T_c$ discussed above for our model parameters, we obtain $w\approx0.97\xi;1.95\xi$, where $\xi^2=\sqrt{D_1D_2}/(2\pi T_c)$, for $D_1/D_2=0.2;10$, respectively. This values coincide with numerics presented in Fig. \ref{f5} remarkably well. 

As expected, the length scales $w_{\Delta_{1,2}}$ obtained in Fig. \ref{f5} are very close due to almost identical spatial profiles of $F_{1,2}$ caused by the efficient interband pairing, see Fig. \ref{f2}A,B. Similarly to the one-band case, scales $w_{\Delta_{1,2}}$
can be fitted by function $(B/H_{c2})^{-1/3}$ in the vicinity of $H_{c2}$, see inset of Fig. \ref{f5}A. However, the characteristic length scales of LDOS modifications, $w_{\sigma_{1,2}}$, demonstrate striking difference with the single-band scenario. Their field dependencies can be both stronger and weaker than that for the gap profiles determined by $w_{\Delta_{1,2}}$. In particular, in the case $D_1\ll D_2$ characterized by the leading role of the vortex-cell radius on the spatial evolution of LDOS in the band with larger diffusion constant (see discussion of Figs. \ref{f2}C and \ref{f4}C) we obtain the stronger field behaviour $w_{\sigma_2}\propto R\propto (B/H_{c2})^{-1/2}$ in the vicinity of $H_{c2}$, see also inset in Fig. \ref{f5}A.
The opposite regime, $D_1\gg D_2$ shown in Fig. \ref{f5}B is described by the exceptionally weak field dependence of LDOS in the band with smaller diffusion coefficient in agreement with discussions of Figs. \ref{f2}D and \ref{f4}D.

\section{Summary}\label{summary}
 To conclude, we demonstrate that the vortex core size $w_{\Delta_k}$ determined by the healing of the gap order parameter has qualitatively different magnetic-field behaviour from the one $w_{\sigma_k}$ defined by the spatial LDOS variations in single- and two-band dirty superconductors.
We have found several generic regimes peculiar for multiband superconductor only. First, the vortex core size $w_{\sigma_k}$ related to the LDOS variations in the band with {\it larger} diffusion constant scales with the vortex-cell radius having field dependence stronger than the one for $w_{\Delta_k}$. Second, size $w_{\sigma_k}$ determined by the LDOS variations in the band with {\it smaller} diffusion constant can have field dependence significantly weaker than for $w_{\Delta_k}$. These peculiarities can  explain qualitatively the recent STM measurements of vortex cores in multiband superconductors.

\section{Acknowledgements}
This work was supported by the Academy of Finland. It is our pleasure to acknowledge discussions with 
   Hermann Suderow and Vladimir Kogan.
\bibliography{bibl}
	
\end{document}